\documentclass[runningheads]{svmult}

\usepackage{makeidx}   
\usepackage{graphicx}  
\usepackage{subeqnar}  
\usepackage{multicol}  
\usepackage{physprbb}  
\makeindex             

\begin{document}

\title*{Optical/near-IR observations of
Gamma-Ray\protect\newline Bursts in the Afterglow Era}

\titlerunning{optical/NIR observations of GRBs in the Afterglow Era}

\author{Alberto J. Castro-Tirado\inst{1,2}}

\authorrunning{A. J. Castro-Tirado}

\institute{Instituto de Astrof\'{\i}sica de Andaluc\'{\i}a (IAA-CSIC), 
P.O. Box 03004, E-18080 Granada, Spain 
\and Laboratorio de Astrof\'{\i}sica Espacial y F\'{\i}sica Fundamental 
     (LAEFF-INTA), P.O. Box 50727, E-28080 Madrid, Spain}

\maketitle              

\begin{abstract}
An overview of the optical and near-IR observations of GRBs
in the Afterglow Era is presented. They have allowed to a better
understanding of the underlying physics as well as to constraint the
progenitor models.  
\end{abstract}

\section{Afterglow observations}

\subsection{Photometric observations}

\subsubsection{Power-law declines.}

{\it BSAX} made possible to detect the first X-ray afterglow following 
GRB 970228 whose precise localization (1$^{\prime}$) led to the discovery 
of the first optical transient (or optical afterglow, OA) associated to
a GRB [1,2].
The OA was afterwards found on earlier images [3,4] and the light curve 
exhibited a power-law (PL)
decay  F $\propto$  t$^{-\alpha}$ with  $\alpha$ =  1.1 [5], thus
confirming the prediction of the relativistic blast-wave model [6]. 
PL declines have been measured for 26 OAs in 1997-2000 yielding 
values in the range  
0.8 $<$  $\alpha$ $<$  2.3 with  $<\alpha>$ = 1.35. 

\subsubsection{Breaks and steepening in light curves.}

A break deviating from a PL decay was first observed in the GRB 990123 
light curve at T$_0$ + 1.5 $d$  ($\sim$ 1.5 days after the high energy event) 
and it was interpreted 
as the presence of a beamed outflow with
a half opening angle $\theta_0$ $\sim$ 0.1  [7-9], i.e. reducing the inferred
energy by a factor of $\sim$ $\theta_0$$^2$/4.
Further breaks have been reported in another 5 GRBs:
GRB 990510 [10-11], GRB 991208 [12], GRB 991216 [13,14],
GRB 000301C [15,16] and GRB 000926 [17,18].

There are six possible explanations for the observed breaks:
i) sideways expansion of the jet caused by the swept-up matter [19,20] 
although the effect might be negligible for $\theta_0$ $>$ 0.1 [21];
ii) when the jet material propagates in an uniform density       
medium and the observer sees the edge of the jet, with $\alpha$ 
increasing by $\sim$ 0.7 [22];
iii) when the jet material propagates in a medium with a PL     
density profile and the observer sees the edge of the jet, $\alpha$ 
increases by a factor of $\Gamma^2$ [22] with $\Gamma$ the bulk Lorentz
factor;
iv) when the jet material propagates within a pre-ejected stellar 
wind ($\rho$ $\propto$  r$^{-2}$)  and the observer sees the edge of 
the jet,  $\alpha$ increases by $\sim$  0.4 [22];
v) when in both the relativistic and non-relativistic cases (if 
$\rho$ is high) the inverse Compton scattering is important, the 
light curves can be flattened or steepened  [21]; and
vi) when the transition from the relativistic phase to the 
non-relativistic phase of an isotropic blastwave takes place in a 
dense medium ($\rho$ $\sim$ 10$^{6}$ cm$^{-3}$) [23].

Rapid fading ($\alpha$ $>$ 2.0) has been observed in 3 GRBs:
GRB 980326 [24], GRB 980519 [25,26] and GRB 991208 [12].
Two possible causes can explain such behaviour: 
the synchrotron emission during the mildly
relativistic and non-relativistic phases [23] and 
the interaction of a spherical burst with a pre-burst
Wolf-Rayet star wind [27,28].

\subsubsection{"Plateau" states.}

For GRB 970508, a "plateau" ($\alpha$ = 0) was observed between 
T$_0$ + 3 $hr$  and  T$_0$ + 1 $d$ [29,30]. 
The optical light  curve
reached a peak in two days [31,32] 
and was followed  by a PL
decay F $\propto$   t$^{-1.2}$ .  The "plateau" has been explained by several
plasmoids with different fluxes occurring at different times [33].
Another "plateau" was detected in the near-IR light curve of GRB
971214 between T$_0$ + 3 $hr$ and T$_0$ + 7 $hr$ [34].

\subsubsection{Short-term variability.}

Flux fluctuations are expected due to inhomogeneities in the     
surrounding medium  as a consequence of 
interstellar turbulence or by variability and anisotropy in a precursor 
wind from the GRB progenitor [35].
However,  short-term variability was found neither in GRB 970508 [30] 
 nor in GRB 990510 [11].       
In GRB 000301C, the high variability observed at optical
wavelengths [14,15] can be due to several reasons:
i) refreshed shock effects;
ii) energy injection by a strongly magnetic millisecond      
pulsar  born during the GRB [36]; 
iii) an ultra-relativistic shock in a dense medium rapidly     
evolving to a non-relativistic phase [37]; and
iv) a gravitational microlens [38].

\subsubsection{The SN-GRB association.}

A peculiar type Ib/c supernova (SN 1998bw) was found in
the error box for the soft GRB 980425 [39] coincident with 
a galaxy at z = 0.0085, but this SN/GRB relationship
is still under debate. 

In any case, "SN-like" bumps have been detected in other GRBs:
GRB 970228 [40,41], GRB 970508 [42], GRB 980326 [43,44],
GRB 980703 [45], GRB 991208 [12] and GRB 000418 [46].
There are some alternative explanations for the existence of such a bump 
in the OA light curves:
i) scattering of a prompt optical burst by 0.1 M$_{\odot}$  dust beyond its
sublimation radius 0.1-1 pc from the burst, producing an echo 
after  20-30 $d$ [47]; 
ii) delayed energy injection by shell collision [48]; and
iii) an axially symmetric jet surrounded by a less energetic
outflow [49]. 
But this is certainly not the case for all GRBs:  GRB 990712 provided the 
first firm evidence that an underlying SN was not present [50].

\begin{figure}[th]
\begin{center}
\includegraphics[width=.7\textwidth]{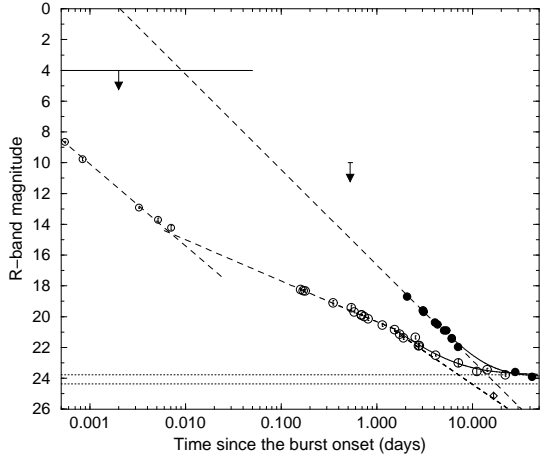}
\end{center}
\caption[]{The two brightest optical GRB afterglows detected 
           so far: GRB 990123 (empty circles) and GRB 991208 (filled circles). 
           The dotted lines are the constant contribution of the two host 
           galaxies, R $\sim$ 23.9 and 24.3 respectively. The dashed-lines 
           are the pure OAs contributions to the total fluxes. The solid 
           lines (only shown here after T $>$ 5 $d$) are the 
           combined fluxes (OA plus underlying galaxy on each case).
           Upper limits are for GRB 991208. Adapted from [12].}
\label{eps1}
\end{figure}

\subsection{Spectroscopy.}

GRB 970508 was the clue to the distance: optical spectroscopy obtained 
during the OA maximum brightness allowed a direct determination of a lower 
limit for the redshift ($z$ $\geq$ 0.835), implying D $\geq$ 4 Gpc 
(for H$_0$ = 65  km s$^{-1}$ Mpc$^{-1}$) and E $\geq$ 7 $\times$ 10$^{51}$ 
erg. 
It was the first proof that GRB sources lie at cosmological distances [51]. 
The flattening of the decay at T$_0$ + 100 $d$ [30,52] 
revealed the contribution of a constant 
brightness source -the host galaxy- seen in late-time imaging 
at T$_0$ + 1 $yr$. 
The 15 GRB redshifts measured so far are in the range
0.430 $\leq$ z $\leq$ 4.50 [53] with $<z>$ = 1.5 and they were derived either
from absorption lines in the OA spectrum, 
from the Ly-$\alpha$ break,  
or from emission lines arising in the host galaxy.

\subsection{Polarimetry.}

As synchrotron radiation under favourable conditions can be
up to 70 \% polarized, the first polarimetric observations were attempted 
for GRB 990123, but only an upper limit was established ($\Pi$ $<$ 2.3\%) [54].
Polarized optical emission was first detected in GRB 990510
 ($\Pi$  = 1.7 $\pm$ 0.2\%) by means of observations
performed at T$_0$ + 18.5 $hr$ [55], T$_0$ + 21 $hr$ and T$_0$ + 43 $hr$ [56].
 This confirmed the synchrotron origin of the blast wave itself and
represented another case for a jet-like outflow [11].
Further polarization measurements were carried out in GRB
990712 during a 1 $d$ time interval (at T$_0$ + 10 $hr$, T$_0$ + 17 $hr$ and
T$_0$ + 35 $hr$). In that case, the polarization angle did not vary 
significantly but the degree of polarization was not constant [18] which 
can be explained by a laterally expanding jet [57]. See also [58,59].

\section{Near-simultaneous GRB observations}

Significant early optical emission may arise from the reverse
shock [27,60], i.e. 
strong optical/near-IR flashes accompanying gamma-ray emission 
should be a generic characteristic (at least for typical GRBs, 
with E $\sim$ 10$^{53}$ erg and $\rho$ $\sim$ 1 cm$^{-3}$ [61]).
Such observations will allow: i) to derive $\Gamma$ by the relative timing 
of optical and  $\gamma$-ray emission, ii) to pinpoint the process by which 
the shells responsible for the external shock arise, and iii) to 
constraint the environment.

ROTSE [62] achieved the detection simultaneously to the GRB of 
the bright optical emission from GRB 990123 [63]:
the most luminous object ever recorded (Fig. 1), with M$_V$ = $-$36  (peaking 
at m$_V$ = 8.9), implying that at least some subsets of GRBs do exhibit 
variable optical emission as violent as the gamma-ray variations.
However, upper limits (in the range R = 4-15) were derived for prompt optical 
emission of a dozen of bursts by means of ROTSE and other experiments, 
like LOTIS [64], TAROT [65], BOOTES [66,67], EON [68] and CONCAM [69]. 
Thus, bright optical counterparts are uncommon.
Why most optical flashes are not detected? Due to several reasons:
i) lack of deeper coverage, given the wide GRB luminosity
function [70];
ii) fireballs in low-density environments ($\rho$ $\ll$ 1 cm$^{-3}$) would
not be expected to produce strong prompt emission;
iii) the reverse shock energy is radiated at a non-optical
frequency, with the synchrotron peak frequency $\nu_m$ $\gg$  $\nu_{opt}$ 
or  $\nu_m$ $\ll$  $\nu_{opt}$ [71]; and
iv) highly absorbed GRB by dust in their host galaxies.

\section{"Dark" GRBs}

The first ``dark'' event (GRB 970828) was detected as a fading X-ray
source [72], although no optical counterpart down to m$_R$ = 24 at T$_0$ 
+ 4 $hr$ was detected [73].
At least in another three cases (GRB 981226, GRB 990506 and GRB 001109), 
radiotransients were detected without accompanying
optical/IR transients. For GRB 000210, the X--ray
position (1$^{\prime\prime}$ accuracy) coincides with a faint galaxy [74]. 

About 40\% of the GRBs with X-ray counterparts do not show OAs, and this 
could be due to: i) intrinsic faintness because of a low ambient medium; 
ii) Lyman limit absorption in high redshift galaxies ($z$ $>$ 7); and 
iii) high absorption in a dusty enviroment: if GRBs are tightly related 
to star-formation, a substantial fraction of them should occur in highly 
obscured regions.

\begin{figure}[th]
\begin{center}
\includegraphics[width=.6\textwidth]{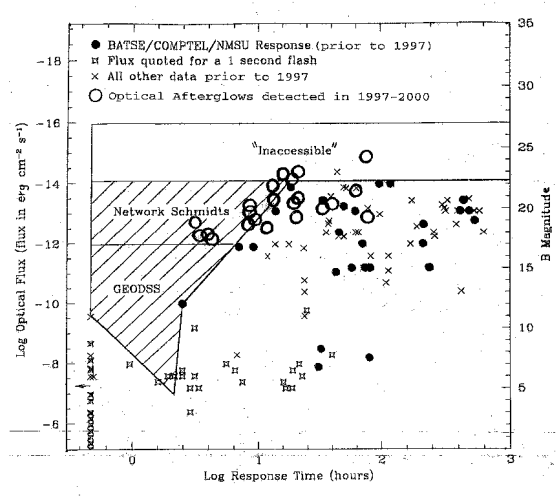}
\end{center}
\caption[]{The GRB follow-up response prior to 1997 (based on [75]) 
compared to the detection of optical counterparts in the 
Afterglow Era (since 1997).
It is clearly seen that the fact the no optical afterglows were discovered
prior to the BSAX launch was just a matter of bad luck (i.e. the $\sim$ 10 
events for which prompt deep follow-up optical searches were performed 
seemed to be either ``rapidly-fading'' or ``dark'' GRBs).}
\label{eps1}
\end{figure}

\section{Summary}

The first optical/near-IR counterparts have been found for $\sim$ 30 precisely 
localized GRBs in 1997-2000 although they
should have been discovered prior to the {\it BSAX} launch (Fig. 2). 
In any case, only the population of GRBs with durations of few
seconds has been explored (see [76] for a more extensive review). 
Short bursts lasting less than 1 s, 
that follow the $-$3/2 slope in the log N-log S diagram (in contrast 
to the longer bursts) remain to be detected at longer wavelengths.
Future missions should be able to address some of
the issues still to be solved, i.e. prompt optical and near-IR observations
should be persued !

%

\end{document}